%% file: perov.tex
\input{macros.tex}
\documentclass[12pt,a4paper,epsfig]{iopart}
\usepackage{epsfig}
\begin{document}
\title[Optical properties of perovskite alkaline earth titanates]{Optical properties of perovskite alkaline earth titanates : a
formulation}
\author{\bf Kamal Krishna Saha, Tanusri Saha-Dasgupta and Abhijit Mookerjee}
\address { S. N. Bose National Centre for Basic Sciences.
Block-JD, Sector-III, Kolkata-700098, India.}
\author{\bf Sonali Saha and T.P. Sinha}
\address{Department of Physics, J.C. Bose Institute, 93/1 Acharya Prafulla Chandra Road,
Kolkata-700009, India}
\begin{abstract}{ In this communication we suggest a formulation of the optical conductivity as a
convolution of an energy resolved joint density of states and an energy-frequency labelled transition rate.
Our final aim is to develop a scheme based on the augmented space recursion for random systems.
In order to gain confidence in our formulation,
we apply the formulation to three alkaline earth titanates $CaTiO_3$,~$SrTiO_3$ and \ba and compare our results with
available data on optical properties of these systems.
 }
\end{abstract}
\jl{3}
\ead{kamal@bose.res.in}
\pacs{71.20,71,20c}
\maketitle
\parindent 0pt
\section{Introduction}

The object of our present study is to derive an expression for the optical
conductivity as a convolution of the energy resolved joint density of states
and an energy-frequency dependent transition rate. The need is to go beyond the
usual reciprocal space based formulations and obtain an expression which we
can immediately generalize for disordered systems. This would require  labelling
states by energy and the  angular momentum labels ($\ell,m$) alone.
Once we derive this expression we shall find a representation
for the optical conductivity in the minimal basis set
of the tight-binding linearized muffin-tin orbitals (TB-LMTO).
The generalization to disordered systems will be carried out through the augmented space
recursion (ASR) introduced by us earlier for the study of electronic properties of
disordered systems \cite{Am}-\cite{Am3}. The ASR carries out the configuration averaging
essential to the description of properties of disordered systems, going beyond the usual
mean-field approaches and taking into account configuration fluctuations.
The input into the ASR method includes the Hamiltonian parameters of the pure constituents,
as the starting point of the local spin density approximation (LSDA) iterations for the alloy. It also includes the information
about the transition rates of the pure constituents, expressed as functions of the
initial and final state energies. The aim of this paper is to reformulate the reciprocal
space representation of the transition rate and re-express it in the energy-frequency label
representation for the pure constituents. Only when we are confident that this works, can we
proceed with the full calculations for the disordered alloy.
 This
communication is an attempt to verify our formulation for a series of alkaline
earth  titanates in the paraelectric phase,
on which extensive theoretical and experimental data of optical properties are available for comparison.

Perovskite structured titanate ferroelectric compounds is, to date, one of the most extensively investigated
materials. They are extremely interesting from the viewpoint of solid state theoreticians because
their structures are a lot simpler than that of any other ferroelectric material known, and, therefore, prove to be
rather simple systems to study and  better understand the ferroelectric phenomenon. The titanates are very
easily prepared as polycrystalline ceramics, they are chemically and mechanically pretty stable and they exhibit
para- to ferroelectric phase transition at or above room temperature.	Continued interest in these compounds has
led to a wide variety of theoretical and experimental work, specially on lattice vibrations. A less common approach has
been those based on electronic structure calculations \cite{Castet}. Michel-Calendini and Mesnard \cite{MM1}-\cite{MM2}
have reported band structure of \ba within a linear combination of atomic orbitals (LCAO) method
with empirical off-diagonal integrals. The pioneering work on \sr was that of Kahn and
Leyendecker \cite{KL}. This was followed by an Augmented Plane Wave (APW)
calculation by Matheiss \cite{M} and a self-consistent tight-binding
calculation by Soules \etal \cite{S}. However, Battaye \etal \cite{Bat} have compared experimental valence-band spectra
with these early theoretical predictions and have concluded that the agreement was not satisfying. Pertosa and Michel-Calendini
\cite{PM} carried out a modified tight-binding calculation on \ba and \sr and compared their results with X-ray photoelectron spectra.
These authors introduced inner orbital interactions.  Perkins and Winter \cite{PW} have carried out LCAO
calculations on the band structure of $SrTiO_3.$  There have been several all-electron, full-potential linearized augmented
plane waves (FP-LAPW) studies of the titanates
in recent times \cite{CK}-\cite{SB}. In addition ultrasoft-pseudopotential, local density approximation (LDA)
based studies on perovskites have been
carried out by King-Smith and Vanderbilt \cite{KSV}. In comparison, electronic structure calculations on \ca have been
fewer. Ueda and coworkers \cite{Ueda1}-\cite{Ueda2} have used the first-principles tight-binding method to study \ca .

 We shall show that for all the three compounds the transition
rate, defined by us, is strongly energy and frequency dependent, i.e. it depends upon	 the energy of both the initial	 and
the final states. We shall compare the theoretical results with experiment.

\section{Methodology}

In recent years a number of methods have been proposed for calculating optical properties within the framework
of the LMTO \cite{Us}-\cite{Ho} for both metals and semi-conductors.
We shall present here a gauge-independent formalism, following the ideas of Hobbs \etal \cite{Ho}. Since our
final aim is to use the augmented space recursion method (ASR) \cite{Am}-\cite{Am3} and study
the optical properties of random systems, we shall modify the reciprocal space
formulation and obtain an expression in	 which all states are labelled by their
energy and the optical conductivity  is expressed as a convolution of the energy resolved joint density of states
and an energy-frequency dependent transition matrix. This formulation will then be directly
generalized within the ASR.

The Hamiltonian describing the effect of a radiation field on the electronic states of a solid is
given by :

\[
H\eq  \sum_{i=1}^{N}\left\{ \frac{1}{2m_e}\left( {\mathbf{p}}_i + {\frac{e}{c} \mathbf{A}}({\mathbf{r}}_i,t)\right)^2 + V({\mathbf{r}}_i)
+ e\Phi(\mathbf{r}_i,t)\right\}\]

Here $e$ is the magnitude of electronic charge, $m_e$ the electronic mass, $c$ is the velocity of light and
$\hbar$ is the Planck's constant.  
${\mathbf{A}}({\mathbf{r}}_i,t)$ and $\Phi(\mathbf{r}_i,t)$ are the vector and scalar potentials seen by the $i$-th electron because of the
radiation field.
 There are N electrons labelled by $i$.
The potential $V({\bf r}_i)$ experienced by the electrons is expressed as an effective independent electron
approximation within the LDA of the density functional theory (DFT). For
not too large external optical fields, neglecting terms of the order of $O(\vert\mbox{\bf A}\vert^2)$, the Hamiltonian
reduces to :

\be
H\eq  \sum_{i=1}^{N}\left\{ \frac{1}{2m_e} {\mathbf{p}}_i^2 \pls  V({\mathbf{r}}_i) \pls \frac{1}{c} \ {\mathbf {j}}_i \cdot {\mathbf {A}}({\mathbf {r}}_i,t) \right\}
\ee

Here {\bf j}$_i$ = (e/m) {\bf p}$_i$ is the current operator.
We work in the Coulomb gauge where  ${\mathbf {\nabla}}\cdot\mathbf{A}({\mathbf{r}}_i,t)$ = 0 and $\Phi({\mathbf{r}}_i,t)$ = 0, so that the electric field

\[ \mathbf{E}({\mathbf{r}}_i,t) =  - \frac{\partial \mathbf{A}({\mathbf{r}}_i,t)}{\partial t}  \]

In choosing the above equation we have ignored the response of the system. The local electric field is the external
field due to the incident radiation as well as the internal field due to the polarization of the medium. Such local
field corrections are important for insulators. We intend, as is customary, to introduce
the local filed corrections as well as corrections due to the Coulomb hole in our
final GW calculations, for which these single-particle picture will form the zeroth
starting point.

The Kubo formula then relates the linear current response to the radiation field  :

\[ \langle j_{\mu}(t) \rangle \eq \sum_{\nu}\int_{-\infty}^{\infty} \ dt'\ \chi_{\mu\nu}(t-t') A_{\nu} (t') \]

The generalized susceptibility is given by :

\[ \chi_{\mu\nu}(\tau) \eq  \imath \Theta(\tau)\ \langle \phi_0\vert [j_{\mu}(\tau),j_{\nu}(0)]\vert \phi_0\rangle \]

where, $\tau=t-t'$ and $\Theta(\tau)$ is the Heaviside step function,

\[ \Theta(\tau) \eq \left\{ \begin{array}{ll}
			     1 & \mbox{if } \tau\ >\ 0 \\
			     0 & \mbox{if } \tau \ \leq \ 0
			    \end{array} \right. \]

$\vert\phi_0\rangle$ is the ground state of the unperturbed system, that is the solid in the absence of the radiation field. In
 the absence of the radiation field, there is no photocurrent, i.e. $\langle\phi_0\vert j_{\mu}\vert\phi_0\rangle$ = $0$.
The fluctuation-dissipation theorem  relates the imaginary part of the generalized susceptibility to the
correlation function as follows :

\be
\chi_{\mu\nu}^{\prime\prime}(\omega) \eq \frac{1}{2} \left(1-e^{-\beta\omega}\right) S_{\mu\nu}(\omega)
\ee

\noindent where,

\[ \beta=\frac{1}{k_BT} \>\>\> \mbox{where $k_B$ is the Boltzmann constant and $T$ the temperature}\]

and 

\[ \chi_{\mu\nu}^{\prime\prime}(\omega)\eq \Im m \int_{-\infty}^{\infty}\ dt\ e^{\imath z \tau}\ \chi_{\mu\nu}(\tau) \quad\quad
\mbox{z=$\omega$+$\imath 0^+$} \]

\noindent and,

\[ S_{\mu\nu}(\omega) \eq \Im m \int_{-\infty}^{\infty}\ dt\ e^{\iota z \tau}\ \langle\phi_0\vert j_{\mu}(\tau)\ j_{\nu}(0)\vert
\phi_0\rangle \quad\quad
\mbox{z=$\omega$+$\imath 0^+$} \]

An expression for the correlation function,  can be obtained via the Kubo-Greenwood expression,

\be S(\omega)\eq \frac{\pi}{3} \sum_{i}\sum_{f} \sum_{\mu} \langle \phi_i \vert j_{\mu} \vert \phi_f\rangle \cdot
\langle \phi_f \vert j_{\mu} \vert \phi_i\rangle \ \delta(E_f - E_i - \hbar\omega)\label{ref1}\ee

We have assumed isotropy of the response so that the tensor $S_{\mu\nu}$ is diagonal and we have defined $S(\omega)$ as the direction averaged quantity
$\frac{1}{3}\sum_{\mu} S_{\mu\mu}(\omega)$. The $\vert\{\phi_i\}\rangle$  are the occupied `initial' single electronic states in the ground state
while $\vert\{\phi_f\}\rangle$ are the unoccupied single electron `final' excited states in the LDA description.

The imaginary part of the dielectric function is
related to the above :

\be
\epsilon_2(\omega)\  =\	 \frac{1}{\pi^2 \omega^2} S(\omega)
\ee

We may obtain the real part of the dielectric function $\epsilon_1(\omega)$ from a Kramers-Kr\"onig
relationship. 

\vskip 0.1cm
For crystalline semi-conductors the equation (\ref{ref1}) may be rewritten as follows :

\be
 S(\omega)\eq \frac{\pi}{3} \sum_{j}\sum_{j'} \int_{BZ} \frac{d^3\mbf{k}}{8\pi^3}\ \left | \langle \Phi_{j'\mbf{k}}\vert\ \mbf{j}
\ \vert \Phi_{j\mbf{k}}\rangle\right |^2 \ \delta\left(E_{j'}(\mbf{k})-E_j(\mbf{k})-\hbar\omega\right)\label{ref2}\ee

Here $j$ and $j'$ refer to band labels : $j$ for the occupied valence bands and $j'$ the unoccupied conduction bands at T=0 K. The
$\mbf{k}$ is the quantum label associated with the Bloch Theorem. For disordered materials, the Bloch Theorem fails
and the expression (\ref{ref2}) can no longer be used. Our first aim will be to obtain an alternative
expression where the quantum states are directly labelled by energy and frequency, rather than by the `band' and `crystal
momentum' indeces.  For this, let us examine the following expressions :

\begin{eqnarray}
 n(E) & \eq & \sum_{j}\int_{BZ} \frac{d^3\mbf{k}}{8\pi^3}\ \delta(E_j(\mbf{k})-E) \label{eq6} \\
\phantom{xx} & & \nonumber \\
 J(E,\omega) & \eq& \sum_{j}^{occ}\sum_{j'}^{unocc}\int_{BZ} \frac{d^3\mbf{k}}{8\pi^3}\ \delta(E_j(\mbf{k})-E)\ \delta(E_{j'}(\mbf{k})
-E-\hbar\omega)\label{eq7}
\end{eqnarray}

In the equation (\ref{eq6}), the right hand side picks up a factor of 1 whenever a quantum state, labelled by $\{ \mbf{k}, j \}$ falls in the range
E, E+$\delta$E. The left-hand side, therefore, is the {\sl density of states} arising from the bands labelled $j$.

In the equation (\ref{eq7}), the right-hand side picks up a factor of 1 whenever a quantum state in the filled bands labelled $j$ falls in the range
E,  E+$\delta$E	 {\sl and simultaneously} a quantum state in the unfilled bands labelled $j'$ falls in the range
E+$\omega$, E+$\omega$+$\delta$E. The left-hand side is then the {\sl energy resolved joint density of states} :

\be J(E,\omega)\ =\ n_v(E)\  n_c(E+\hbar\omega) \ee

We shall define the energy-frequency labelled {\sl Transition rate} as :

\vskip 0.1cm
\begin{center}
 $T(E,\omega)$ \eq \begin{tabular}{c}{$\displaystyle{\sum_{j}\sum_{j'}\int_{BZ} \frac{d^3\mbf{k}}{8\pi^3}\ T^{jj'}(\mbf{k})
\ \delta(E_j(\mbf{k})-E)\ \delta(E_{j'}(\mbf{k})-E-\hbar\omega)}$}\\ \hline
{$\displaystyle{\sum_{j}\sum_{j'}\int_{BZ} \frac{d^3\mbf{k}}{8\pi^3}\ \delta(E_j(\mbf{k})-E)\ \delta(E_{j'}(\mbf{k})-E-\hbar\omega)}$}
\end{tabular}
\end{center}
\be \phantom{xxx} \label{ref4}\ee

Where,
\[ T^{jj'}(\mbf{k})\eq \sum_{\mu}\left | \rule{0mm}{4mm} \langle \Phi_{j'\mbf{k}}\vert\ j_{\mu}
\ \vert \Phi_{j\mbf{k}}\rangle\right |^2 \]

The expression for $S(\omega)$ from equation (\ref{ref2}) then becomes,

\be
S(\omega) \eq (\pi/3)\ \int\ dE\ T(E,\omega)\ J(E,\omega)
\label{ref3}
\ee

Many earlier workers argued that the transition matrix element is weakly dependent on both E and $\omega$. They then
assumed it to be constant $T_0$ and obtained a simple expression for the correlation function :

\be S_0(\omega) \eq (\pi/3)\ T_0 \int\ dE\ J(E,\omega) \ee

We shall investigate the validity of this approximation for the systems under study in this communication.
 Let us first
get an expression for the equation (\ref{ref3}) within the TB-LMTO formalism of \cite{Ander1}-\cite{Skriver} : The basis of the
LMTO starts from the minimal muffin-tin orbital basis set of a KKR formalism and then linearizes it by expanding
around a `nodal' energy point $E_{\nu \ell}^\alpha$~. The wave-function is then expanded in this basis :

\begin{eqnarray}
\Phi_{j\mbf{k}}(\mbf{r}) & \eq & \sum_{L}\sum_{\alpha} c^{j\mbf{k}}_{L\alpha}\left[ \phi_{\nu L}^\alpha
(\mbf{r})+\sum_{L'}\sum_{\alpha '}\ h_{LL'}^{\alpha\alpha '}(\mbf{k})\ {\mathaccent 95 \phi}_{\nu L'}
^{\alpha '}(\mbf{r})\right]\nonumber
\end{eqnarray}

\noindent where, $L$ is the composite angular momentum	index $(\ell,m)$, $j$ is the band index and $\alpha$ labels
the atom in the unit cell.

\noindent and,
\begin{eqnarray*}
\phi_{\nu L}^\alpha({\mbf{r}}) & = & \imath^{\ell}\ Y_L(\hat{r})\ \phi_\ell^\alpha(r,E_{\nu\ell}^\alpha) \\
{\mathaccent 95 \phi}_{\nu L}^\alpha(\mbf{r}) & = & \rule{0mm}{6mm}\imath^{\ell}\ Y_L(\hat{r})\ \frac{\partial
\phi_\ell^\alpha(r,E_{\nu \ell}^\alpha)}{\partial E} \\
h_{LL'}^{\alpha\alpha '}(\mbf{k}) & = & (C_L^\alpha - E_{\nu \ell}^\alpha)\ \delta_{LL'}\delta_{\alpha\alpha '} \pls \sqrt{\Delta_{L}^\alpha}\ S_{LL'}^{\alpha\alpha '}(\mbf{k})\ \sqrt{\Delta_{L'}^{\alpha '}} \\
\end{eqnarray*}

 $C_L^\alpha$ and $\Delta_L^\alpha$ are TB-LMTO potential parameters and  $S_{LL'}^{\alpha\alpha '}(\mbf{k})$ is the structure matrix. These terms are standard for the LMTO formulation and the reader is referred to the citation  \cite{Ander1}
for greater detail. The TB-LMTO secular
equation provides the  expansion coefficients $c^{j\mbf{k}}_{L\alpha}$ via :

\begin{equation} \sum_{L'}\sum_{\alpha '} \left[ h_{LL'}^{\alpha\alpha '}(\mbf{k}) + (E_{\nu\ell}^\alpha \mns E^{j\mbf{k}})\ \delta_{LL'}\ \delta_{\alpha\alpha '} \right]\  c^{j\mbf{k}}_{L'\alpha '} \eq 0 \label{sec}\end{equation}

We may now immediately write an expression for the matrix element of the current operator as in equation (\ref{ref4}) :

\begin{eqnarray}
\fl \langle \Phi_{j'\mbf{k}}(\mbf{r}) \vert\ \mbf{j}\ \vert \Phi_{j\mbf{k}}(\mbf{r})\rangle  \eq
 \sum_{LL'}\sum_{\alpha} \bar{c}^{j'\mbf{k}}_{L'\alpha}\ c^{j\mbf{k}}_{L\alpha}
\left\{ \rule{0mm}{6mm}\langle \phi_{\nu L'}^{\alpha}(\mbf{r})\vert\ \mbf{j}\
\vert \phi_{\nu L}^\alpha (\mbf{r})\rangle  \ldots\right. \nonumber \\
  \left. \ldots\pls \rule{0mm}{6mm}(E^{j\mbf{k}}-E_{\nu\ell}^\alpha)\ \langle \phi_{\nu L^\prime}^{\alpha}
(\mbf{r})\vert\ \mbf{j}\ \vert {\mathaccent 95 \phi}_{\nu L}^\alpha(\mbf{r})\rangle  \ldots\right.\nonumber\\
 \left.\ldots \pls \rule{0mm}{6mm}(E^{j'\mbf{k}}-E_{\nu\ell'}^{\alpha})\ \langle {\mathaccent 95 \phi}_{\nu L^\prime}
^{\alpha}(\mbf{r})\vert\ \mbf{j}\ \vert \phi_{\nu L}^\alpha (\mbf{r})\rangle  \ldots\right. \nonumber \\
  \left.\ldots\pls (E^{j\mbf{k}}-E_{\nu\ell}^\alpha)\ (E^{j'\mbf{k}}-E_{\nu\ell'}^{\alpha})\ \langle {\mathaccent 95 \phi}_{\nu L^\prime}^{\alpha}(\mbf{r})\vert\ \mbf{j}\ \vert {\mathaccent 95 \phi_{\nu L}^\alpha}(\mbf{r})\rangle  \rule{0mm}{6mm}\right\} \nonumber\\
\end{eqnarray}

We shall now obtain expressions for the right-hand terms by  noting the following :

\begin{eqnarray}
\mbf{j} \eq e \ \rule{0mm}{7mm} \frac{d\mbf{r}}{dt} & \eq & \frac{e}{\imath\hbar} [\mbf{r}, H]\nonumber \\
\rule{0mm}{6mm} H \phi_{\nu L}^\alpha(\mbf{r}) & \eq &	E_{\nu\ell}^\alpha\ \phi_{\nu L}^\alpha(\mbf{r})\nonumber \\
 \rule{0mm}{6mm} H {\mathaccent 95 \phi}_{\nu L}^\alpha(\mbf{r})& \eq & \phi_{\nu L}^\alpha(\mbf{r}) \pls E_{\nu\ell}^\alpha \ {\mathaccent 95 \phi}_{\nu L}^\alpha(\mbf{r})
\end{eqnarray}

We can write,
\[  \mbf{r} \eq (2\pi/3)^{1/2}\ r \left[ (Y_{1,-1}-Y_{1,1})\ \hat{\mbf{i}} \pls	 \imath (Y_{1,-1}+Y_{1,1})\ \hat{\mbf{j}}
\pls 2^{1/2} Y_{1,0}\ \hat{\mbf{k}}\right] \]

Using the above two equations we get,

\be
\int {\phi^{\alpha}_{\nu L'}(\mbf{r})}^{\star}\ \mbf{r}H\ \phi_{\nu L}^\alpha (\mbf{r})\ d^3{\mbf{r}} \eq  \imath^{\ell -\ell '}\ E_{\nu\ell}^\alpha\ \mbf{\Gamma}_{LL'}\ \int_{0}^{s_\alpha}
\phi_{\nu\ell '}^{\alpha}(r)\ \phi_{\nu\ell}^\alpha (r)\  r^3 dr
\ee

where, $s_\alpha$ is the atomic sphere radius of the $\alpha$-th atom in the unit cell
and ${\mbf{\Gamma}}_{LL'}$ is a combination of Gaunt coefficients \cite{Ho} :

\[
\fl {\mbf{\Gamma}}_{LL'} \eq \sqrt{\frac{2\pi}{3}}\left[\ \left(G^{m',-1,m}_{\ell ',1,\ell} - G^{m',1,m}_{\ell ',1,\ell}\right) \hat{\mbf{i}} \pls
  \imath \left( G^{m',-1,m}_{\ell ',1,\ell} + G^{m',1,m}_{\ell ',1,\ell}\right) \hat{\mbf{j}} \pls
  \sqrt{2}\ G^{m',0,m}_{\ell ',1,\ell} \hat{\mbf{k}}\ \right]
\]

In order to obtain $\int {\phi^{\alpha}_{\nu L'}(\mbf{r})}^{\star}\ H \mbf{r}\ \phi_{\nu L}^{\alpha}(\mbf{r})\ d^3\mbf{r}$, we note that $H\eq (\hbar^2/2m_e)\mbf{\nabla}^2 + V(\mbf{r})$
so that using the Green's  second identity  we can obtain,

\begin{eqnarray}
 \int {\phi^{\alpha}_{\nu L'}(\mbf{r})}^{\star}\ H\ \mbf{r}\ \phi_{\nu L}^\alpha (\mbf{r})\ d^3\mbf{r}	 \eq
\imath^{\ell -\ell '} \mbf{\Gamma}_{LL'}\left\{ E_{\nu\ell '}^\alpha \int_{0}^{s_\alpha} \phi_{\nu\ell '}^\alpha (r) \ \phi_{\nu\ell}^\alpha (r) r^3 dr
 \ldots \right.\nonumber \\
\rule{0mm}{6mm} \phantom{xxxxxxxxxx}\ldots  \left.  \pls (\hbar^2/2m_e)s^2_\alpha \ \phi_{\nu\ell}^\alpha (s_\alpha)\ \phi_{\nu\ell '}^\alpha (s_\alpha)\	 \left( D_{\nu\ell '}^\alpha - D_{\nu\ell}^\alpha -1\right)\right\} \nonumber \\
\end{eqnarray}

$D_{\nu\ell}^\alpha$ is the logarithmic derivative of $\phi_{\nu\ell}^\alpha (r)$ at  $r\eq s_\alpha$ and are obtained as parameters in the
TB-LMTO routines. We
 define the following integrals~:

\begin{eqnarray*}
\int_{0}^{s_\alpha}\ \phi_{\nu\ell'}^\alpha (r)\ \phi_{\nu\ell}^\alpha (r)\ r^3 dr & \eq & I_{\ell\ell'}^\alpha \\
\int_{0}^{s_\alpha}\ \phi_{\nu\ell'}^\alpha (r)\ {\mathaccent 95 \phi}_{\nu\ell}^\alpha (r)\ r^3 dr & \eq & J_{\ell\ell'}^\alpha \\
\int_{0}^{s_\alpha}\ {\mathaccent 95 \phi}_{\nu\ell'}^\alpha (r)\ {\mathaccent 95 \phi}_{\nu\ell}^\alpha (r)\ r^3 dr & \eq & K_{\ell\ell'}^\alpha
\end{eqnarray*}

call,
\[ (\hbar^2/2m_e)s_{\alpha}^2(D_{\nu\ell}^\alpha -D_{\nu\ell'}^\alpha -1) \eq  {\mathcal D}_{\ell\ell '}^\alpha \]

Then the matrix elements for the current operator becomes :

\begin{eqnarray}
\fl {\mathcal I}^{(1)\ \alpha}_{LL',\mu}\eq \langle \phi_{\nu L'}^\alpha (\mbf{r})\vert\ j_{\mu}\ \vert\phi_{\nu L}^\alpha (\mbf{r})\rangle & \eq &\frac{\imath^{\ell -\ell'-1}}{\hbar} \Gamma_{LL'}^\mu
\left[ \tilde{E} \ I_{\ell'\ell}^\alpha - {\mathcal D}_{\ell '\ell}^\alpha\ \phi_{\nu\ell}^\alpha (s_\alpha)\ \phi_{\nu\ell'}^\alpha (s_\alpha)\right] \nonumber \\
\fl{\mathcal I}^{(2)\ \alpha}_{LL',\mu}\eq\langle \phi_{\nu L'}^\alpha (\mbf{r})\vert\ j_\mu\ \vert{\mathaccent 95 \phi}_{\nu L}^\alpha (\mbf{r})\rangle& \eq &\frac{\imath^{\ell -\ell'-1}}{\hbar} \Gamma_{LL'}^\mu
\left[ \tilde{E} \ J_{\ell'\ell}^\alpha + I_{\ell'\ell}^\alpha	- {\mathcal D}_{\ell '\ell}^\alpha \ {\mathaccent 95 \phi}_{\nu\ell}^\alpha (s_\alpha )\ \phi_{\nu\ell'}^\alpha (s_\alpha)
\right] \nonumber \\
\fl {\mathcal I}^{(3)\ \alpha}_{LL',\mu}\eq\langle {\mathaccent 95 \phi}_{\nu L'}^\alpha (\mbf{r})\vert\ j_\mu\ \vert \phi_{\nu L}^\alpha (\mbf{r})\rangle & \eq &\frac{\imath^{\ell -\ell'-1}}{\hbar} \Gamma_{LL'}^\mu
\left[ \tilde{E} \ J_{\ell\ell'}^\alpha - I_{\ell'\ell}^\alpha	- {\mathcal D}_{\ell '\ell}^\alpha\ \phi_{\nu\ell}^\alpha (s_\alpha )\ {\mathaccent 95 \phi}_{\nu\ell'}^\alpha (s_\alpha ) \right]\nonumber \\
\fl {\mathcal I}^{(4)\ \alpha }_{LL',\mu}\eq\langle {\mathaccent 95 \phi}_{\nu L'}^\alpha (\mbf{r})\vert\ j_\mu\ \vert {\mathaccent 95 \phi}_{\nu L}^\alpha (\mbf{r})\rangle & \eq &
\frac{\imath^{\ell -\ell'-1}}{\hbar} \Gamma_{LL'}^\mu
\left[\rule{0mm}{5mm} \tilde{E} \ K_{\ell\ell'}^\alpha + J_{\ell'\ell}^\alpha -J_{\ell\ell'}^\alpha  \ldots\right.\nonumber\\
& & \left.\phantom{xxxxxxxxx}\ldots - {\mathcal D}_{\ell '\ell}^\alpha \ {\mathaccent 95 \phi}_{\nu\ell}^\alpha (s_\alpha)\ {\mathaccent 95 \phi}_{\nu\ell'}^\alpha (s_\alpha)\rule{0mm}{5mm}\right] \nonumber \\
\phantom{x}\nonumber\\
\end{eqnarray}

where, $ \tilde{E} = E_{\nu\ell}^\alpha - E_{\nu\ell'}^\alpha $

The transition term $T^{jj'}({\bf k})$ has to be written in terms of the normalized wavefunction.

The normalizing factor for the wavefunctions are obtained from :

\begin{eqnarray}
N^{j}_{\mbf{k}} & \eq  & \int\ d^{3}\mbf{r}\ \Phi_{j\mbf{k}}^{\star}(\mbf{r})\ \Phi_{j\mbf{k}}(\mbf{r}) \nonumber\\
		& \eq & \sum_{L}\sum_{\alpha} \vert\bar{c}_{L\alpha }^{j\mbf{k}}\vert^2\ \left\{{\mathcal J}^{(1)}_{L\alpha} + 2(E^{j\mbf{k}}-E_{\nu\ell}^\alpha)
			 {\mathcal J}^{(2)}_{L\alpha} + (E^{j\mbf{k}}-E_{\nu\ell}^\alpha)^2
			    {\mathcal J}^{(3)}_{L\alpha }\right\} \nonumber\\
\end{eqnarray}
where,
\begin{eqnarray*}
{\mathcal J}^{(1)}_{L\alpha} &\eq & \int_{0}^{s_\alpha} \left| \phi_{\nu L}^\alpha (r) \right|^2\ r^2dr \\
{\mathcal J}^{(2)}_{L\alpha} & \eq & \int_{0}^{s_\alpha} {\phi^{\alpha}_{\nu L}(r)}^{\star}\ {\mathaccent 95 \phi}_{\nu L}^\alpha (r)\ r^2dr \\
{\mathcal J}^{(3)}_{L\alpha} & \eq & \int_{0}^{s_\alpha} \left| {\mathaccent 95 \phi}_{\nu L}^\alpha (r)\right|^2\ r^2dr \\
\end{eqnarray*}

Using the secular equation (\ref{sec}), the expression for the transition term becomes,
\begin{eqnarray}
\fl  T^{jj'}(\mbf{k})  \eq
 \left( N^{j}_{\mbf{k}}N^{j'}_{\mbf{k}}\right)^{-1/2}\left |\rule{0mm}{6mm} \sum_{\mu}\sum_{LL'}\sum_{\alpha}
\bar{c}^{j'\mbf{k}}_{L'\alpha}c^{j\mbf{k}}_{L\alpha}\left\{ {\mathcal I}^{(1)\ \alpha}_{LL',\mu} +
  (E^{j\mbf{k}}-E_{\nu\ell }^\alpha)\ {\mathcal I}^{(2)\ \alpha}_{LL',\mu} \ldots \right.\right.\nonumber \\
\fl   \phantom{xxxxxxxxx} \left.\left.\ldots + (E^{j'\mbf{k}}-E_{\nu \ell'}^\alpha) {\mathcal I}^{(3)\ \alpha}_{LL',\mu}
 + (E^{j\mbf{k}}-E_{\nu\ell }^\alpha)(E^{j'\mbf{k}}-E_{\nu \ell'}^\alpha) {\mathcal I}^{(4)\ \alpha}_{LL',\mu}
   \right\}\rule{0mm}{6mm} \right|^{2}\nonumber\\
\phantom{x}\nonumber\\
\end{eqnarray}

The equation (10) provides an expression for the optical conductivity where both the transition matrix and the
energy resolved joint density of states are expressed as functions of energy and frequency. As we shall show in
a subsequent communication, that within the ASR formalism, this is the form in which the information
about the constituents are input and the configuration averaged correlation function for the alloy may
be expressed as :

\be \ll S(\omega)\gg \eq (\pi/3)\ \int\ dE\ T^{eff}(E,\omega)\ \ll J(E,\omega)\gg \ee

where,

\[ \fl \ll J(E,\omega)\gg \eq  \ll n_v(E)\gg \ll n_c(E+\omega)\gg \left[ \rule{0mm}{4mm}1\pls \Lambda(E,\omega) \ll J(E,\omega)\gg \right] \]

and,

\[ T^{eff}(E,\omega)\eq \ll T(E,\omega)\gg \pls \delta T\left( \rule{0mm}{4mm} E,\omega,\Sigma(E,\omega)\right) \]

where, $\Sigma\left(E,\omega\right)$ is the self-energy due to disorder scattering and $\Lambda\left(E,\omega\right)$ the corresponding
vertex correction. The details of the derivation will be communicated in a subsequent paper \cite{op2}.

\section{Calculational Details and Results}

\begin{figure}[t]
\centering
\epsfxsize=5.5 in \epsfysize=2.1 in \rotatebox{0}{\epsfbox{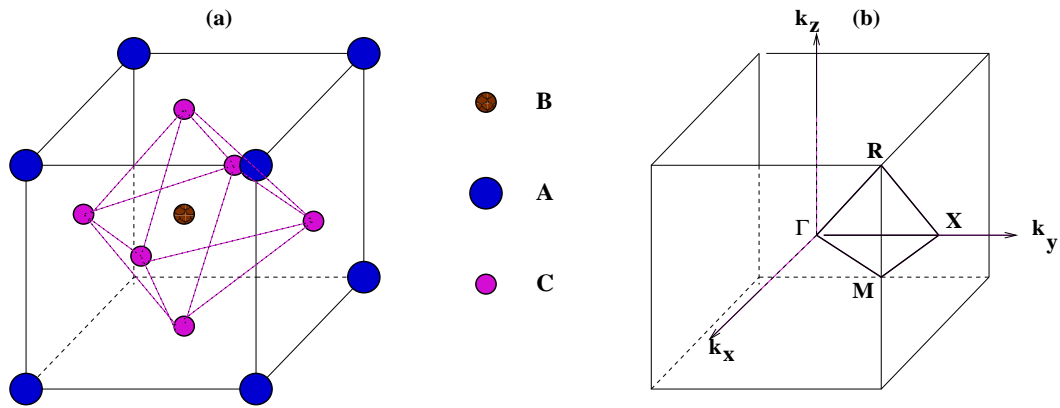}}
\caption{(a) Cubic unit cell for a perovskite $ABC_3$.
(b) The Brilluoin zone for the cubic phase}
\label{fig1}
\end{figure}

The primitive cell for the ideal perovskite structure $ABC_3$ is illustrated in figure
\ref{fig1} (a). For the class of compounds we are interested in, the generic chemical
formula is $ABO_3$. $A$	 is a mono or divalent cation, $B$ is a tetra or
penta-valent metal. In the paraelectric phase there is full cubic symmetry. It
can be thought of as lattice of corner sharing oxygen octahedra with inter-penetrating
simple cubic lattices of $A$ and $B$. The $B$ cations sit in the centre of the octahedral
O cage, while the $A$ metal ions sit in the 12-fold coordinated sites between the octahedra.
In our case the body centre position is occupied by the Ti atom, the edges by
alkaline earth atoms and the face centres by O atoms. The space group is $O^{1}_h$ and the
corresponding Brilluoin zone is shown in figure~\ref{fig1} (b). Both the A and B atoms
are situated at sites with full cubic ($O_h$) point symmetry, while the O atoms have
tetragonal ($D_{4h}$) symmetry. 

The electronic configuration of the alkali earth atoms are as follows :

\begin{center}
\begin{tabular}{lllll}\br
Atom & Deep & Shallow & Valence & Unoccupied \\ 
$\>$ & Core & Core & $\>$ & $\>$ \\ \mr
Ca & $1s^2\ 2s^2\ 2p^6\ 3s^2\ 3p^6$					 & -	  & $4s^2$	    & $3d\ 4p$ \\
Sr &  $1s^2\ 2s^2\ 2p^6\ 3s^2\ 3p^6\ 3d^{10}\ 4s^2$			 & $4p^6$ & $5s^2$	    & $4d\ 5p\ 4f$ \\
Ba &  $1s^2\ 2s^2\ 2p^6\ 3s^2\ 3p^6\ 3d^{10}\ 4s^2\ 4p^6\ 4d^{10}\ 5s^2$ & $5p^6$ & $6s^2$	    & $5d\ 6p\ 4f$ \\
Ti & $1s^2\ 2s^2\ 2p^6\ 3s^2\ 3p^6 $					 & -	  & $3d^2\ 4s^2$ & $4p$ \\
O  & $1s^2$								 & -	  & $2s^2\ 2p^4$    & $3s\ 3d$ \\ \br
\end{tabular}
\end{center}

Since we wish to take into account the shallow core states, to include the transitions from these to the conduction band
at large enough optical frequencies, the energy range is about 40 eV (3 Ryd) and the single panel
LMTO cannot be made to be accurate over this range, we have carried out a two panel calculation, with the
$E_{\nu\ell}^{\alpha}$ lying in the lower energy range in one panel and in the upper energy range in the other.
The minimal basis set used in the two panels are the following :

\begin{center}\begin{tabular}{ll}\br
\multicolumn{2}{c}{\ca } \\ \mr
Lower Panel  &	Ca $4s\ 4p\ 3d$, Ti $4s\ 4p\ 3d$, O $2s\ 2p\ (3d)$ \\
Upper Panel  &	Ca $4s\ 4p\ 3d$, Ti $4s\ 4p\ 3d$, O $3s\ 2p\ (3d)$ \\ \br
\multicolumn{2}{c}{\sr } \\ \mr
Lower Panel  &	Sr $5s\ 4p\ 4d\ (4f)$, Ti $4s\ 4p\ 3d$, O $2s\ 2p\ (3d)$ \\
Upper Panel  &	Sr $5s\ 5p\ 4d\ (4f)$, Ti $4s\ 4p\ 3d$, O $3s\ 2p\ (3d)$ \\ \br
\multicolumn{2}{c}{\ba } \\ \mr
Lower Panel  &	Ba $6s\ 5p\ 5d\ (4f)$, Ti $4s\ 4p\ 3d$, O $2s\ 2p\ (3d)$ \\
Upper Panel  &	Ba $6s\ 6p\ 3d\ (4f)$, Ti $4s\ 4p\ 3d$, O $3s\ 2p\ (3d)$ \\ \br
\end{tabular}\end{center}

\begin{figure}[h]
\centering
\epsfxsize=5.6 in \epsfysize=3.8 in \epsfbox{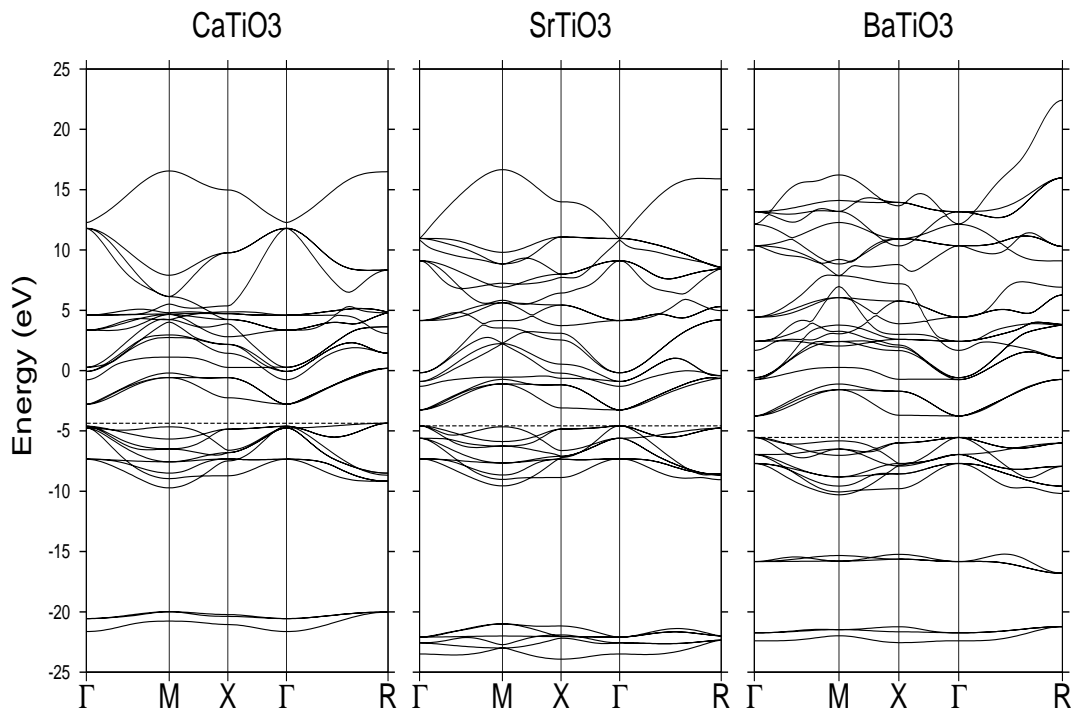}
\caption{Band structures of \ca,\ \sr and \ba (No scissors operation has been carried out in these
calculations)}
\label{fig2}
\end{figure}

The states in parentheses are unfilled states which have been downfolded in our
calculations. The figure \ref{fig2} shows the band structure of the three titanates.

Let us first look at the leftmost figure for \ca.
 The Ca $3p$ core level lies around
--27 eV and not shown in this figure. The narrow band around --20 eV is the O $2s$ band. The nine valence
bands just below the Fermi level are derived from hybridized Ti $4s$ and O $2p$. An indirect band gap
appears between the valence band top at the R point and the conduction band minimum at the $\Gamma$ point.
\ca shows an indirect
band gap ($\Gamma$-R) of 1.6 eV, in the absence of the {\sl scissors} operation. The experimentally
reported indirect gap is 3.5 eV \cite{Ueda1}. This discrepancy is characteristic of the Local
density Approximation (LDA) upon which the TB-LMTO is based. In the conduction band region we have
bands originating from (in ascending order of energy) Ti $3d\ t_{2g}$ triplet, a singlet arising from 
Ca $4s$  and  a doublet from Ti $3d\ e_g$ . Then comes the bands which originate from the  Ca $3d\ e_g$ 
doublet and the Ca $3d\ t_{2g}$ triplet.
 Finally we have the Ti $4p$ and Ca $4p$  based bands and finally the band based on Ca $4s$. We note that for \ca, Ca
and Ti $3d$ based bands overlap and hybridize in the conduction region.

For \sr, in the lower panel, the Sr $4p$ level now sits almost atop the O $2s$ band, giving rise to a
rather broad (as compared to \ca) $s-p$ hybridized band just below --20 eV. The subsequent analysis of the bands
is rather similar to \ca. However, the band gap is now direct and $\sim$ 1.4 eV. Earlier band structure calculations
of Mo \etal \cite{Mo} and Kimura \etal \cite{Kim} yield indirect band gaps of $\sim$ 1.45 eV and $\sim$ 1.79 eV
respectively. The experimental direct band gap is around 3.2 eV.

For \ba, in the lower panel, the Ba $5p$ shallow core level now crosses and lies above the O  $2s$ band. The band gap is
direct and $\sim$ 1.2 eV. The band structure is almost identical to the pseudopotential  calculations of
King-Smith  and Vanderbilt \cite{KSV}, whose band gap was also direct and $\sim$ 1.8 eV. The experimental band
gap turns out to be $\sim$ 3.2 eV \cite{Wem}.

For all three compounds, our calculations show a characteristic flatness of the lowest conduction band along
$\Gamma$ to X. This agrees with earlier works of Cardona \cite{Car}, Matheiss \cite{M}, Harrison \cite{H},
Wolfram and Ellialtio\u{g}lu \cite{WE} and King-Smith and Vanderbilt \cite{KSV}. This observed flatness is related to certain unusual features in the density of states and optical conductivity, which appear to be characteristic of pseudo-
two-dimensional systems.

\begin{figure}
\centering
\epsfxsize=4 in \epsfysize=5 in \rotatebox{0}{\epsfbox{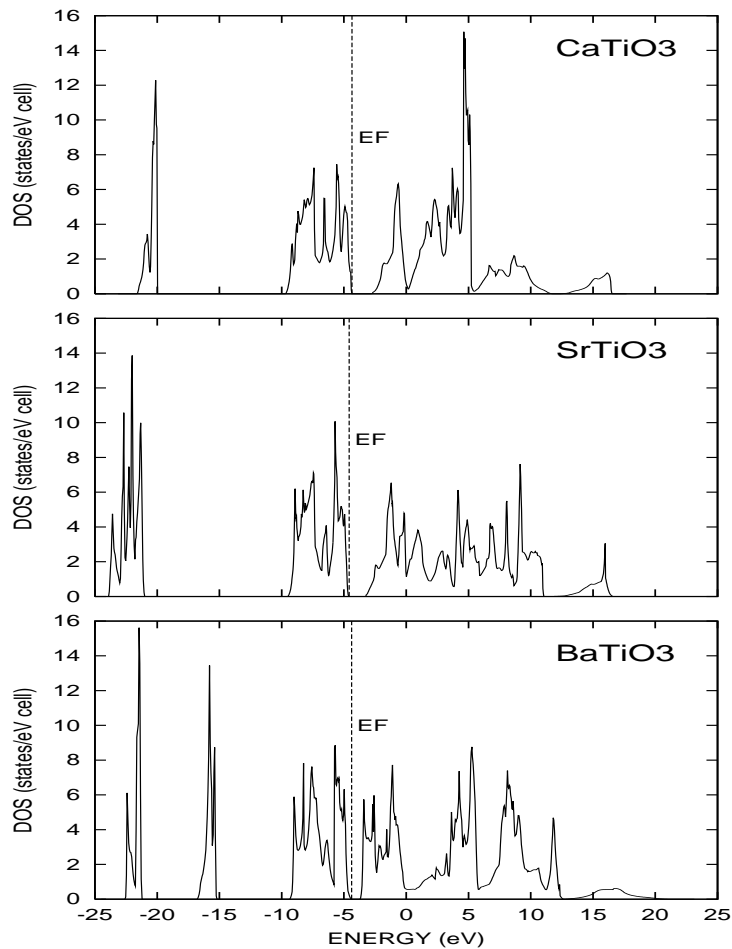}}
\caption{Densities of states for the three perovskite titanates}
\label{fig3}
\end{figure}

Figure \ref{fig3} shows the densities of states for the three titanates. The densities of states reflect the detailed
band structure we have described above. Earlier works on the density of states were based on different
methods. Michel-Calendini and Mesnard \cite{MM1,MM2} and Pertosa and Michel-Calendini \cite{PM} have used
parameterized tight-binding and adjusted LCAO based methods for \ba.  Their density of states is in  agreement with ours, with
the band gap difference characteristic of LSDA methods like the TB-LMTO. Similarly, Matheiss \cite{M} used 
the LCAO and Perkins and Winter \cite{PW} used the extended H\'ukel basis for \sr. Again their band structures
and densities of states are in agreement with ours, with the exception of the band gap. Even earlier works on \sr
by Zook and Castelman \cite{ZC} and Soules \etal \cite{SKVR} also show reasonable agreement. For \ca we may
compare our results with those of Ueda \etal \cite{Ueda1}, which agree reasonably well with our results. 

\begin{figure}[t]
\centering
\epsfxsize=3.0 in \epsfysize=2.3 in \rotatebox{0}{\epsfbox{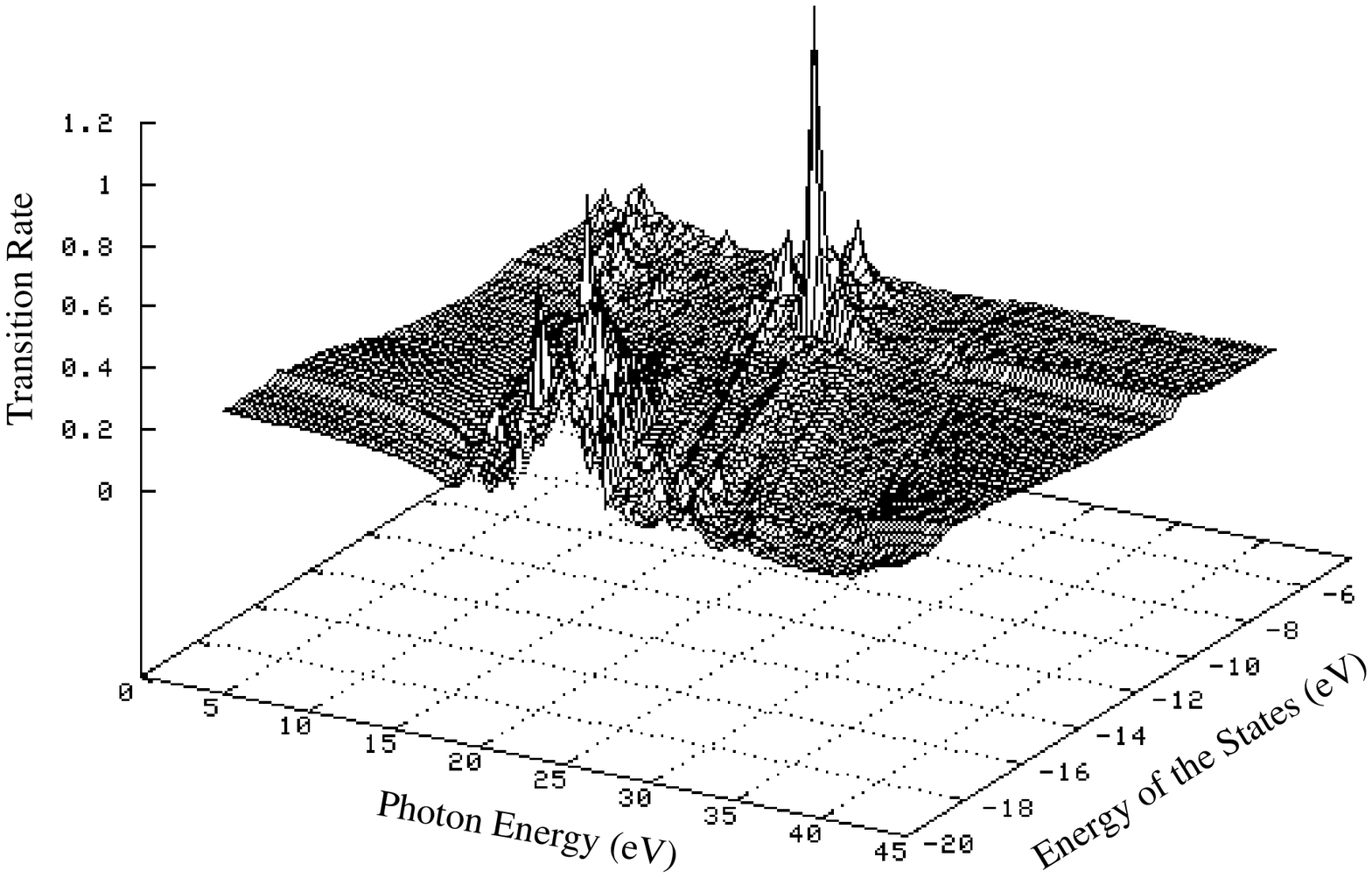}}
\vskip -0.6cm
\epsfxsize=5.8 in \epsfysize=2.2 in \rotatebox{0}{\epsfbox{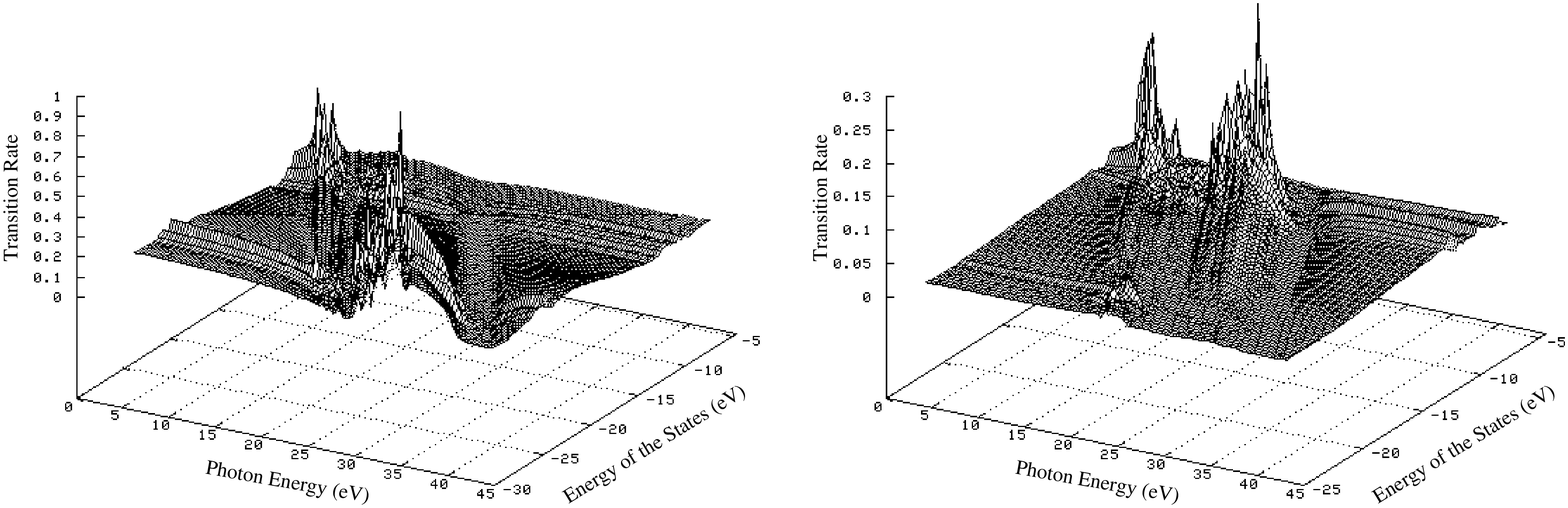}}
\caption{Transition rates for the three perovskite titanates shown as functions
of initial energies and incident photon energies. \ba (top), \sr (left-bottom) and \ca (right-bottom)}
\label{fig4}
\end{figure}

\begin{figure}
\centering
\epsfxsize=5.8 in \epsfysize=4.9 in \epsfbox{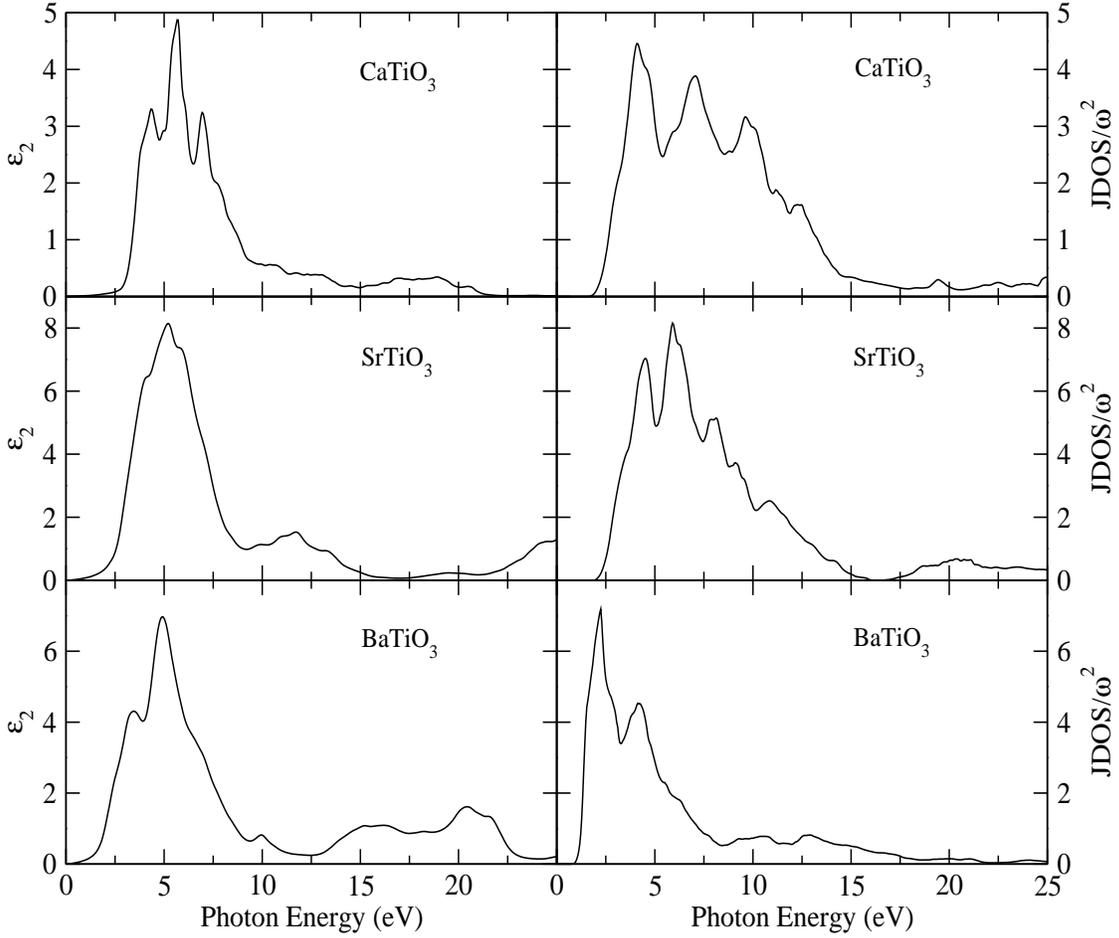}
\caption{Comparison of the calculated imaginary part of the dielectric function
$(\epsilon_2)$ (left) for the three perovskite titanates with the same function 
calculated considering transition rate is constant (right).}
\label{fig5}
\end{figure}

Figure \ref{fig4} displays the transition rates $T(E,\omega)$ for the three titanates, shown here as
functions of the initial energy of the excited electron and the incident photon energy (frequency).
It is clear from the figure that the transition rate is strongly dependent on the energy-frequency
variables for all the three compounds. The usual assumption of a transition matrix, weakly dependent
on energy and frequency is certainly not valid in any of the three cases \cite{SSM1}-\cite{SSM3}.

In figure \ref{fig5} we compare the imaginary part of the dielectric function with 
the scaled joint-density of states/$\omega^2$. If the transition rate were independent
of energy and frequency, they should be the same. The behaviour of the two are
similar, but the relative weights of the structures across the frequency range are
clear indications of the energy-frequency dependence of the transition rates.
\vskip 0.5cm
{\sl \ca } :\\

\begin{figure}
\centering
\epsfxsize=4.5 in \epsfysize=4 in \epsfbox{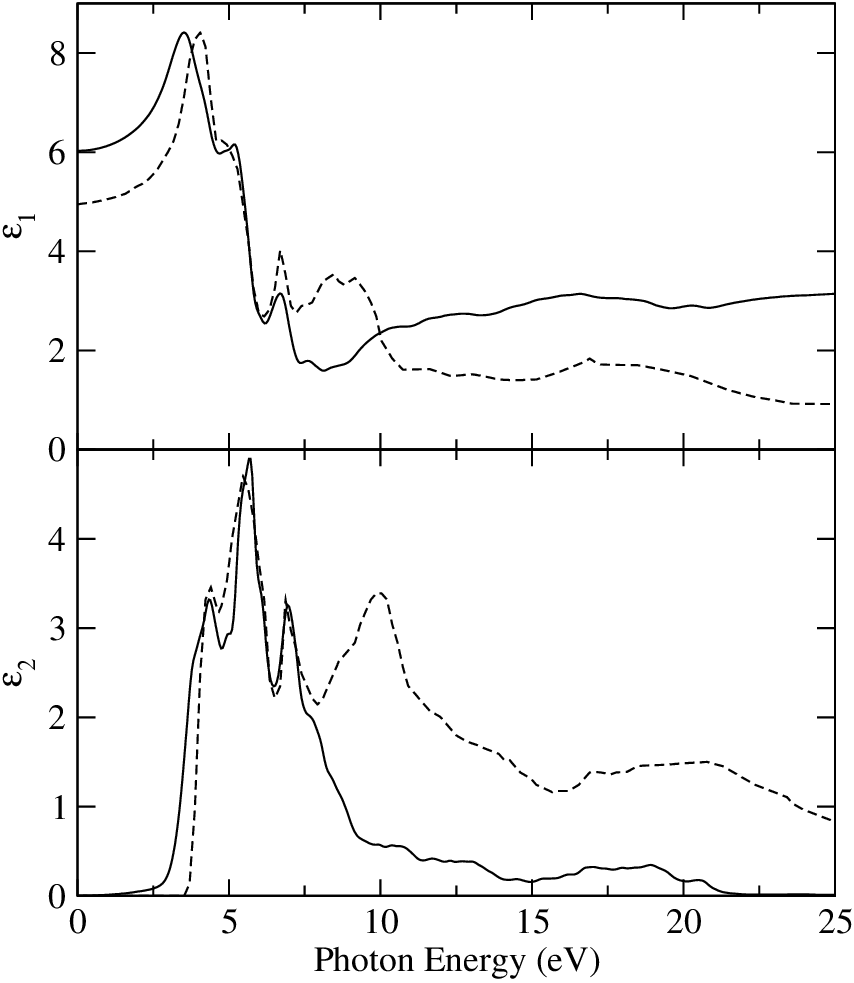}
\caption{Comparison of calculated and experimental real part $\epsilon_1
(\omega)$(top) and imaginary part $\epsilon_2(\omega)$ (bottom) of dielectric
function of \ca as a function of Photon energy: dotted line is
experimental (\cite{Ueda1}-\cite{Ueda2}), continuous line is theoretical.}
\label{fig6}
\end{figure}

The effect of energy-frequency dependence of the transition rate has a large effect for the optical
properties of \ca. If we compare the results reported in \cite{SSM3} with our figure~\ref{fig6} we note
that although in the earlier work the joint density of states does reproduce the peaks at lower frequencies, the relative heights
are not replicated. Looking at the lower panel of the figure,  we may assign the peak at 4.5 eV to the	transitions :
O $2p$ $\rightarrow$ Ti $3d$-$t_{2g}$  at the R	 and X-points. The next and highest peak arises because of
the nearby two unresolved peaks due to the transitions : O $2p$ $\rightarrow$ Ca $4s$ at the M-point (at
6.3 eV) and O $2p$ $\rightarrow$ Ca $3d$-$e_g$ at the M-point (at 6.8 eV). The third peak at 7.5 eV may be assigned
to the transitions : O $2p$ $\rightarrow$ Ti $3d$-$t_{2g}$ at the R-point and O $2p$ $\rightarrow $Ca $3d$-$e_g$
at the X-point. At higher frequencies, the theory does not tally well with experiment. The peak at 10 eV is not
reproduced except as a shoulder. The real part of the dielectric function $\epsilon_1(\omega)$ is obtained
by a Kramers-Kr\"onig transformation from the imaginary part. This is shown in the top panel of
figure \ref{fig6}. 
The discrepancies at high frequencies could be due
to the fact that in the scissors type approach we have provided a rigid shift to the conduction bands. In a full fledged
many-body GW technique, which  is our ultimate aim to produce, the shift, due to the self-energy will turn out to be
energy (frequency) dependent.

\vskip 0.5cm

{\sl \sr } :\\

\begin{figure}
\centering
\epsfxsize=4.5 in \epsfysize=4 in \epsfbox{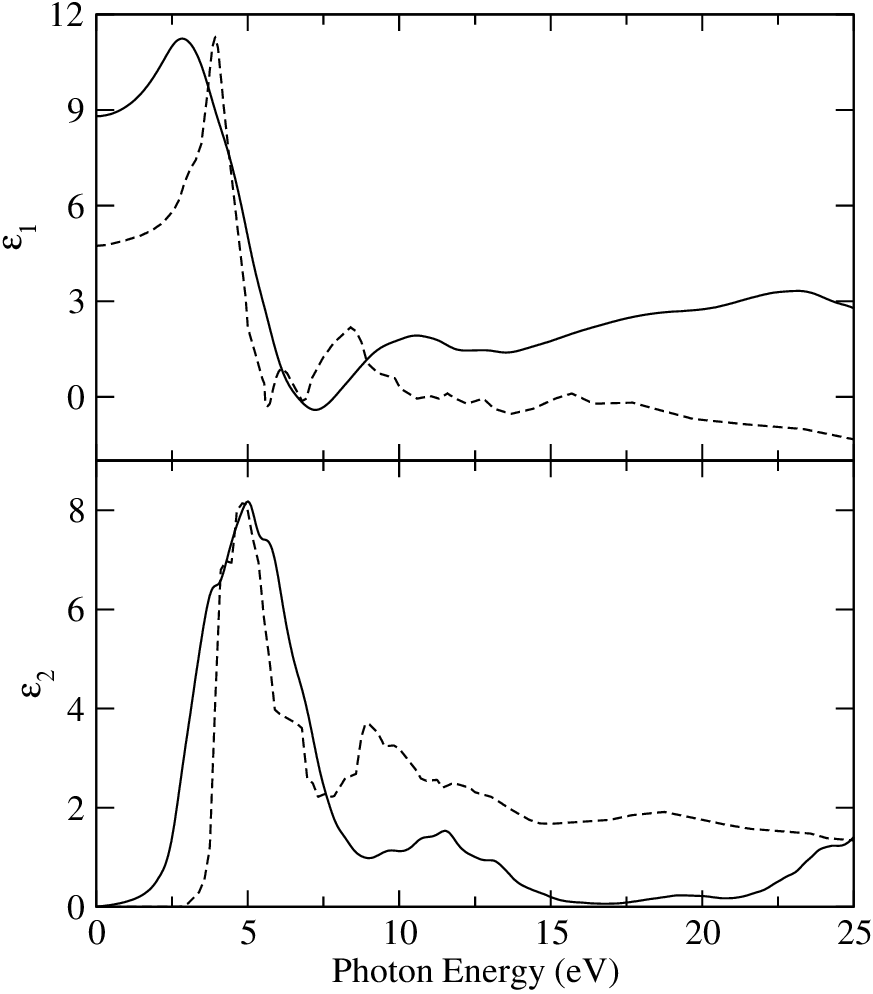}
\caption{Comparison of calculated and experimental real part $\epsilon_1
(\omega)$(top) and imaginary part $\epsilon_2(\omega)$ (bottom) of dielectric
function of \sr as a function of Photon energy: dotted line is
experimental (\cite{Bau}), continuous line is theoretical.}
\label{fig7}
\end{figure}

Let us now examine the figure \ref{fig7}. As in the case of \ca, here too, the effect of energy-frequency dependence of the transition rate reproduces
the correct relative heights of the peaks in $\epsilon_2(\omega)$. The shoulder peak at around 4 eV
may be attributed to the transitions :	O $2p$ at  $\rightarrow$ Ti $3d-t_{2g}$
and O $2p$  $\rightarrow$ Ti $3d-e_{g}$ both at the $\Gamma$ point. The high peak at 5 eV due to O $2p$
$\rightarrow$ Ti $3d$-$e_g$ at the $\Gamma$ point. A third shoulder peak at 6 eV is due to the transition O $2p$
 $\rightarrow$ Ti $3d$-$e_g$ also at the $\Gamma$ point. As in the case of \ca, the structure in the high frequency part
has both lower heights and are shifted to higher frequencies. The cause is the same as discussed above.

\vskip 0.5cm

{\sl \ba }:\\

\begin{figure}
\centering
\epsfxsize=4.5 in \epsfysize=4 in \epsfbox{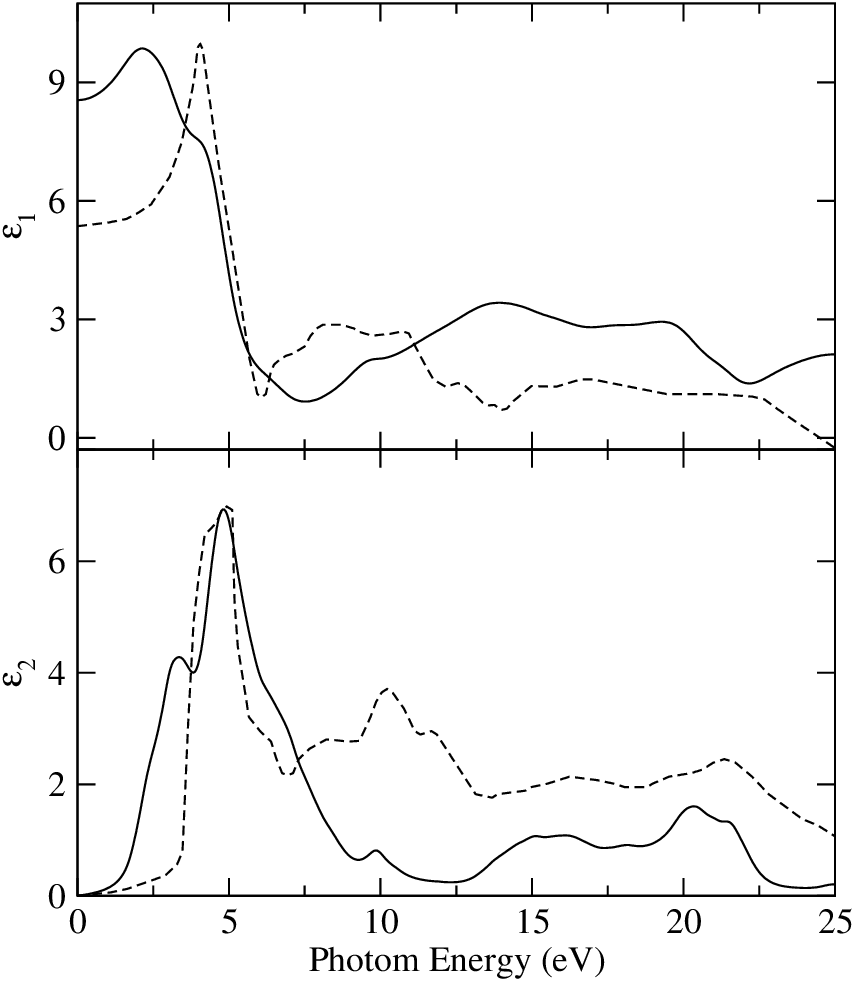}
\caption{Comparison of Calculated and experimental real part $\epsilon_1
(\omega)$(top) and imaginary part $\epsilon_2(\omega)$ (bottom) of dielectric
function of $BaTiO_3$ as a function of Photon energy: dotted line is
experimental (\cite{Bau}), continuous line is theoretical.}
\label{fig8}
\end{figure}

Lastly, let us look at figures \ref{fig6}-\ref{fig8}.
If we compare the shape of the imaginary part of the dielectric function $\epsilon_2(\omega)$ obtained by
our accurate estimate of the transition rate with that of \cite{SSM3}, we note that agreement with experiment \cite{Bau} is
much better when we take the energy-frequency dependences of the transition matrix into account. The relative weights
 of the low frequency peaks, at 3.8 eV and 5 eV are correctly reproduced here. The lower peak is attributed to the
transition from the O $2p$  to	the Ti $3d$-$t_{2g}$  band   and from the O $2p$
 to the Ti $3d$-$e_g$  band both at the $\Gamma$ points. The next higher peak is the because of the transition
from the O $2p$	 to  the Ti $3d$-$e_g$	band at the $\Gamma$ point. Our present study also indicates a
peak around 10 eV and the features at higher frequencies follow the experimental results closely, although the amplitude
seems to have been underestimated as compared to the low frequency results. In all cases, although the lower
frequency part is much better reproduced, the high frequency structures in the theoretical result is shifted upwards.  As before,
we argue that this is probably an artifact of the rigid shift of the scissors type approach. A energy-frequency dependent
self-energy of the type given by the GW method should provide the necessary correction. In addition, the
wrong heights at higher frequencies could also arise from the fact that we have neglected transitions to some of
the higher energy conduction bands. A more complete, perhaps a three panel calculation at higher energies, should
correct this. Alternatively, we could use the higher order NMTOs \cite{kn:tsoka}, when they are available, since they span a much
larger energy range.

\section{Conclusion}

We have proposed here a modified expression for the optical conductivity as a convolution of an 
energy-frequency dependent transition matrix and the energy resolved joint density of states. The main motivation
was to generalize it to disordered systems, where the traditional reciprocal space formulation breaks down
due to the failure of Bloch's Theorem. In order to be confident in our new formulation we have applied it here
to the three alkaline earth perovskite titanates in their paraelectric phases. The results are in
reasonable agreement with experimental data. The agreement is as good as we can expect from a LDA calculation.
 This formulation
 will now form the starting point of a two-fold generalization : first, combining with the ASR to
random systems and as then as a starting point for a many-body GW formulation.

\end{document}

%% file: macros.tex
\def\mbf#1{{\mathbf {#1}}}
\def\eq{\ =\ }
\def\mns{\ -\ }
\def\pls{\ +\ }
\def\be{\begin{equation}}
\def\ee{\end{equation}}
\def\ba{$BaTiO_3$ }
\def\ca{$CaTiO_3$ }
\def\sr{$SrTiO_3$ }